\documentclass[final,3p,times,twocolumn,sort&compress]{elsarticle}

\usepackage{amsmath,amssymb,graphicx,color,psfrag,textcomp}

\newcommand\diff{\mathrm{d}}
\renewcommand\vec[1]{\boldsymbol{\mathrm{#1}}}
\newcommand\op[1]{\mathsf{#1}}
\DeclareMathOperator\Real{Re}

\usepackage[normalem]{ulem}

\journal{Chemical Physics}

\begin{document}

\begin{frontmatter}



\title{Persistent memory for a Brownian walker in a random array of obstacles}


 \author[erlangen,lmu]{Thomas Franosch}
\author[oxford]{Felix H\"of{}ling}
\author[lmu]{Teresa Bauer}
\author[lmu]{Erwin Frey}
\address[erlangen]{Institut f\"ur Theoretische Physik, Universit\"at Erlangen-N\"urnberg, Staudtstra{\ss}e 7, 91058 Erlangen, Germany}
\address[lmu]{Arnold Sommerfeld Center for Theoretical Physics and Center for NanoScience (CeNS), Fakult{\"a}t f{\"u}r Physik,
 Ludwig-Maximilians-Universit{\"a}t M{\"u}nchen, Theresienstra{\ss}e 37, 80333 M{\"u}nchen, Germany}
\address[oxford]{Rudolf Peierls Centre for Theoretical Physics, 1~Keble Road, Oxford OX1 3NP, England, United Kingdom}
\begin{abstract}
\noindent
We show that for particles performing Brownian motion in a frozen array of scatterers long-time correlations
emerge in the mean-square displacement.  Defining the
velocity autocorrelation function (VACF) via the second time-derivative of the mean-square displacement, power-law tails
 govern the long-time dynamics similar to the case of  ballistic motion.
The physical origin of the persistent memory is due to repeated encounters with the same obstacle which occurs
naturally in Brownian dynamics without involving other scattering centers. This observation suggests that in this case
the VACF exhibits these anomalies already at first order in the scattering density. Here we provide an analytic solution for
the dynamics of a tracer for a dilute planar Lorentz gas and compare our results to computer simulations. Our result support
 the idea that quenched disorder provides a generic mechanism for persistent correlations irrespective of
 the microdynamics of the tracer particle.
\end{abstract}


\end{frontmatter}



\section{Introduction}

\noindent
The essence of the Brownian motion of mesoscopic particles suspended in a solvent   has been  understood since
the pioneering works of Einstein~\cite{Einstein:1905} and von Smoluchowski~\cite{vonSmoluchowski:1906}.
Today, Brownian motion constitutes a basic paradigm of stochastic processes~\cite{Haenggi:2005,Haenggi:1982,Frey:2005} with applications in numerous fields,
which may even utilize the noise as by stochastic resonance~\cite{Gammaitoni:1998}
and Brownian motors~\cite{Haenggi:2009,Reimann:2002,Astumian:2002},
not forgetting important generalizations beyond classical statistical physics to quantum Brownian system~\cite{Haenggi:2005b} and relativistic Brownian motion~\cite{Dunkel:2009}.

Following Einstein and von Smoluchowski, the statistics  of the particle's trajectories is described in terms of a probability distribution for the displacement
as was measured shortly after by Perrin~\cite{Perrin:1909} relying on the newly developed dark field microscope.
These displacements are cumulative, i.e., they are the sum of many small steps, and provided the steps are uncorrelated,
the central limit theorem applies.
Then for an unbiased random walk the  probability cloud is described by a Gaussian which broadens as time proceeds.
The corresponding trajectories are continuous but almost nowhere differentiable  curves, self-similar in a statistical sense.
This picture obviously
cannot apply to the very small time and length scales where collisions with individual solvent molecules are resolved. Nevertheless
 it seems plausible that on a coarse-grained time and
length scale all correlations due to these molecular processes are rapidly suppressed and Brownian motion emerges as mathematically
 described by a stochastic differential equation by Langevin~\cite{Langevin:1908} with a Wiener process as noise term.

This established picture had to be reconsidered when Alder and Wainwright~\cite{Alder:1967,Alder:1970} showed in molecular
 dynamics simulations that the basic assumption
of weakly correlated displacements is not fulfilled for the single particle motion in a fluid. Rather they found a long-time
 anomaly $B t^{-d/2}, B>0 $ in the velocity autocorrelation function (VACF) due to
the conservation of momentum in the fluid, where $d$ is the spatial dimension of the system.  For a mesoscopic colloidal
 particle the direct experimental observation of these anomalies {\`a} la Perrin was achieved only 100 years after Einstein's
 work using high-precision photonic force microscopy~\cite{Lukic:2005}. The power law correlations persist even in the presence
 of a bounding wall where momentum can be transferred to,  although they decay more rapidly~\cite{Felderhof:2005,Jeney:2008,Franosch:2009}.

In this paper we argue that quenched disorder constitutes a  second generic mechanism for persistent correlations in the velocity autocorrelation function.
 It was shown in the late 60s that
in the Lorentz model, where a single tracer scatters ballistically in a random array of frozen hard obstacles, ring collisions
 lead to a nonanalytic dependence of the diffusion constant
on the density of scatterers~\cite{Weijland:1968}. The same mechanism implies a power-law tail in the VACF~\cite{Ernst:1971a} of the
 form $- A t^{(d+2)/2}$, $A>0$.
First, in contrast to Alder's discoveries, the origin of this tail is not connected to momentum conservation,
since the tracer exchanges momentum with the frozen scatterers. Second,
the persistent correlation is negative reflecting the caging due to the obstacles. Here the correlations in the velocity
 are inherited from the configuration space since, loosely speaking, the particle remembers forever which paths are blocked by obstacles~\cite{vanBeijeren:1982}.
The predicted anomalies were soon partially confirmed by molecular dynamics simulations~\cite{Bruin:1972,Bruin:1974}
yet the prefactor of the tail was by an order of magnitude larger than expected. At higher densities the relaxation appears
 to become slower suggesting a density-dependent exponent~\cite{Alder:1978,Alder:1983}. Later it was suggested that a second
 universal power law takes over as the the percolation threshold is approached~\cite{Lowe:1993}
in qualitative agreement with a theoretical prediction obtained in a mode-coupling approach~\cite{Goetze:1981a,Goetze:1981b}.
The scenario of a crossover behavior explaining the enhanced prefactor of the power law
was fully confirmed  only recently~\cite{Lorentz_LTT:2007}.

Since the kinetic theory developed  by Weijland and van Leeuwen~\cite{Weijland:1968} is rather involved, one is
 tempted to  assume that their result constitutes a peculiarity of the Lorentz model
with few general consequences. In this paper we show that the long-time anomalies in the VACF persist in a system where
 particles move diffusively rather than ballistically through
the course of obstacles. We calculate analytically the first systematic correction  for the complete scattering
 function and specialize to low wavenumbers to obtain the VACF. The
long-time tail emerges again due to repeated encounters with the same scatterer, yet the derivation of this
 result drastically simplifies compared to the ballistic case, since a diffusive tracer finds the same obstacle
 many times without requiring the series of Boltzmann
scattering events as the ballistic particle. We corroborate our analytic result by Brownian dynamics simulations
 in the dilute regime and discuss the range of validity of the low-density expansion. Last we conclude that
 frozen disorder generically entails long-time algebraic decay
in the VACF irrespective of the microscopic dynamics implying that the basic assumption of quickly decaying
correlations in Brownian motion is generally not fulfilled in heterogeneous media.

The notion of universal long-time anomalies also applies to  hopping transport in disordered lattice models \cite{Nieuwenhuizen:1987,Ernst:1987} where again
repeated encounters with the same scatterer lead to persistently correlated motion.
There the VACF has been calculated up to second order in the obstacle density and the predictions have been nicely confirmed by computer simulation~\cite{Frenkel:1987}.

The paper is organized as follows. In Sec.~\ref{sec:Ballistic}, we briefly recall the main results of the anomalies in the low-density
 expansion of the ballistic Lorentz model.
For the Brownian particle exploring a course of obstacles introduced in Sec.~\ref{sec:multiple}, a multiple scattering expansion is derived
in Sec.~\ref{sec:formal} revealing
 that for the first-order density expansion it is sufficient to solve the single obstacle scattering problem. Section~\ref{sec:single}
calculates  the corresponding forward scattering amplitude which is used in Sec.~\ref{sec:vacf} to discuss the intermediate scattering
function and, in particular, the velocity autocorrelation function in the dilute case. The analytic predictions
 are tested against computer simulation in Sec.~\ref{sec:simulation}.
A summary of the results and conclusions are given in Sec.~\ref{sec:Conclusion}.

\section{Ballistic Lorentz model and kinetic theory}\label{sec:Ballistic}
\noindent
In the Lorentz model  randomly distributed, possibly overlapping obstacles of radius $\sigma$ represent a random,
frozen environment in which a single, point-like
tracer particle moves. The dynamics of the tracer is considered as ballistic with elastic scattering whenever an
 obstacle is encountered. In particular, energy is conserved and the particle's velocity remains constant in magnitude for all times. Diffusion emerges after many collisions with the scatterer as the direction of the velocity is randomized. Since the obstacles are distributed randomly and independently, the structures are characterized solely via the obstacle density $n$. The only dimensionless control parameter is the
reduced obstacle density $n^* := n \sigma^d$.

To lowest order in the obstacle density the motion of a ballistic particle in the course of obstacles is described
 by the Lorentz-Boltzmann equation~\cite{Balian:2007}. There uncorrelated scattering events lead to diffusion on scales larger than the mean-free path $\ell \sim \sigma/n^*$. In this geometric problem the time scale is set by the mean collision time $\tau = \ell/v$ by means of the velocity $v$ of the particle.
The corresponding
diffusion coefficient $D_0 \sim v \ell \sim v \sigma/ n^*$ diverges for small densities since it is due to the rare scattering events only that the ballistic motion
becomes diffusive after all. Via a series of these uncorrelated scattering events the particle may return
 to an already visited obstacle, giving rise to non-trivial correlations. To account for these repeated collision events, one has to go systematically
beyond the Lorentz-Boltzmann equation resulting in a non-analytic correction to the diffusion coefficient $D$. In two dimensions,
$d=2$, to which we restrict the discussion in the following, one obtains ~\cite{Weijland:1968,Bruin:1972},
\begin{equation}
\frac{D_0}{D} = 1 - \frac{4 n^*}{3} \ln n^* -0.8775 n^*+4.519 (n^* \ln n^*)^2 + \ldots
\end{equation}
where the Lorentz-Boltzmann diffusion coefficient is given by $D_0 = 3 v \sigma/16 n^*$.   Uncorrelated scattering events predict an exponential decay
in the
velocity autocorrelation function  $Z(t) := \langle \vec{v}(t) \cdot \vec{v}(0) \rangle/2 = (v^2/2) \exp(-4 t/3 \tau)$ with the mean collision rate $\tau^{-1} = 2 n^* v/\sigma$. The repeated scattering  from the same obstacle then introduces an algebraic  tail
\begin{equation}
 Z(t) \simeq - \frac{\sigma^2}{8 \pi n^*} \frac{1}{t^2}\, , \qquad \text{for } t\to\infty\, , n^*\to 0\, .
\end{equation}
reflecting the infinite memory of the motion in the disordered system. The amplitude diverges as the system becomes more and more dilute, yet since the initial value of the velocity autocorrelation function is density independent, $Z(t=0) = v^2$, the time required to attain the tail is set by the growing collision time $\tau$. Provided time is measured in terms of
$\tau$, the tail is a small $\mathcal{O}(n^*)$ correction to the Lorentz-Boltzmann theory.

In summary, in the framework of the ballistic Lorentz model the non-analytic dependence of $D$ on $n^*$ and the long-time anomaly emerge in a higher order correction
beyond the Lorentz-Boltzmann theory.

\section{Brownian particle in a disordered array}\label{sec:multiple}

\noindent
The Lorentz model can be extended to particles that perform Brownian motion in the void space, i.e the domain not excluded by the frozen scatterers.
In the planar case, this might be a useful minimal model for the complex transport found in  cellular membranes or models thereof~\cite{Dix:2008,Avidin:2010,Sung:2006,Sung:2008a}.
The configuration of the environment is described by the centers of the obstacles $\vec{x}_1,\ldots, \vec{x}_N$ in a finite hypercubic box of length $L$. The density of the scatterers
is kept fixed, $n= N/L^d$, as the thermodynamic limit is performed. Due to exclusion, the distance of the tracer to any of the obstacles always exceeds the radius of the scatterers, $|\vec{r}-\vec{x}_i|\geq \sigma$, $i=1,\ldots,N$. Periodic boundary conditions are assumed throughout for convenience.

A complete statistical description is
given in terms of the conditional probability density $P(\vec{r},t| \vec{R} t')$ to find the particle at time $t$ at position $\vec{r}$ provided
 it was known to be at $\vec{R}$ at an earlier time $t'\leq t$.
Since the stochastic process is stationary, the conditional probability is time-translationally invariant and
we choose $t'=0$ in the following. The Smoluchowski equation for $P(\vec{r},t| \vec{R} 0)$ is then simply the diffusion equation restricted to the void space
\begin{equation}
 \partial_t P(\vec{r},t| \vec{R} 0) = D_0 \nabla^2 P(\vec{r},t| \vec{R} 0)
\end{equation}
where the nabla operator $\vec{\nabla}$ acts on the current position of the particle $\vec{r}$ and now $D_0$ refers to the short-time diffusion coefficient.
Since the tracer particle cannot penetrate the obstacles, this equation of motion has to be supplemented
by the von Neumann boundary condition
\begin{equation}
 (\vec{r}-\vec{x}_i) \cdot \vec{\nabla} P = 0 \, , \qquad \text{for } |\vec{r}-\vec{x}_i| = \sigma, \quad i=1,\ldots,N
\end{equation}
i.e., the flux through the boundary vanishes.

To avoid difficulties with hard core repulsion, we consider for the moment a random potential consisting of  finite spherical barriers,
\begin{equation}
 U(\vec{r}) = \sum_{i=1}^N u(\vec{r}-\vec{x}_i)
\end{equation}
 where, e.g., $u(\vec{r}) = U_0 \vartheta(\sigma-|\vec{r}|)$ with the Heaviside step function $\vartheta(\cdot)$ , and the hard core limit $U_0\to \infty$ is anticipated.
Then the \emph{propagator} obeys the Smoluchowski equation
\begin{equation}
 \partial_t P = \frac{D_0}{k_B T}  \vec{\nabla} \cdot \left(P \vec{\nabla} U \right) + D_0 \nabla^2 P
\end{equation}
where $k_B T$ is the thermal energy.

Analytic progress is made by studying the one-sided temporal Fourier transform
\begin{equation}
 G(\omega; \vec{r}, \vec{R}) = \int_0^\infty \text{e}^{\text{i} \omega t} \, P(\vec{r} t | \vec{R}0) \diff t
\end{equation}
for complex frequencies $\omega$ in the upper half-plane,
rather than studying the time dependence of the propagator $P(\vec{r}t| \vec{R} 0)$ directly.
 Then the equation of motion translates to
\begin{equation}\label{eq:Smoluchowski_frequency}
 (-\text{i} \omega - D_0 \nabla^2) G  - \frac{D_0}{k_B T}  \vec{\nabla} \cdot \left(G \vec{\nabla} U \right)=  \delta(\vec{r}-\vec{R}) \, .
\end{equation}
This form of the Smoluchowski equation identifies $G$ as an inverse of the Smoluchowski operator and constitutes
 the starting point for the elaborated framework of the scattering theory.

\section{Formal scattering theory}\label{sec:formal}

\noindent
The Smoluchowski equation in the form of Eq.~\eqref{eq:Smoluchowski_frequency} has a mathematical analogy
 to the time-independent Schr\"odinger equation, allowing us to employ the techniques developed for the scattering problem of
a quantum particle. First it is convenient to adopt an abstract bra-ket  notation. The Hilbert space is spanned
 by the (generalized) ket states $|\vec{r} \rangle$ normalized by $\langle \vec{r} | \vec{r}' \rangle
= \delta(\vec{r}-\vec{r}')$. Then we introduce an  operator $\op{G}(\omega)$ with
a positional representation $\langle \vec{r} | \op{G}(\omega) | \vec{R} \rangle = G(\omega;\vec{r},\vec{R})$.
The dependence on the complex frequency $\omega$ will be suppressed in
the following. Similarly, we introduce the unperturbed Smoluchowski operator $\op{\Omega}_0$ via its matrix elements
 $\langle \vec{r} | \op{\Omega}_0 | \psi \rangle = D_0 \nabla^2
\langle \vec{r} | \psi \rangle$ and perturbation $\op{V}$ by $\langle \vec{r} | \op{V} | \psi \rangle =
 (D_0/k_B T)  \vec{\nabla} \cdot \left( \langle\vec{r}|\psi\rangle \vec{\nabla} U \right)  $.
The Smoluchowski equation (\ref{eq:Smoluchowski_frequency}) is equivalent to the following operator equation
\begin{equation}
 (-\text{i} \omega - \op{\Omega} ) \op{G} = \mathbb{I} \, ,
\end{equation}
where $\Omega = \Omega_0 + \op{V}$ and $\mathbb{I}$ is the identity operator. The problem is to evaluate
the full Green function $\op{G}$ in the presence of the scatterers. The case of no obstacles leads to the unperturbed Green function $\op{G}_0$ which satisfies
\begin{equation}
 (-\text{i} \omega - \op{\Omega}_0) \op{G}_0 = \mathbb{I} \, .
\end{equation}
Simple operator algebra reveals that the full Green function obeys the Lippmann-Schwinger equation
\begin{equation}
 \op{G} = \op{G}_0 + \op{G}_0 \op{V} \op{G} \, .
\end{equation}
which yields upon iteration the Born series with the first order approximation $\op{G} = \op{G}_0 + \op{G}_0 \op{V} \op{G}_0$.
 The \emph{scattering matrix} defined by $\op{S} := (\mathbb{I} - \op{G}_0 \op{V})^{-1}$  advances solutions of the
 unperturbed problem to the full one $\op{G} = \op{S} \op{G}_0$. It is convenient to single  out the event of no
 scattering and to define the \emph{T-matrix} as $\op{G}_0 \op{T} = \op{S}-\mathbb{I}$. Then the following identities are easily derived
\begin{align}
\label{eq:full_propagator}\op{G} &= \op{G}_0 + \op{G}_0 \op{T} \op{G}_0 \, ,\\
\label{eq:definition_tmatrix} \op{T} &= \op{V} + \op{V} \op{G}_0 \op{T} \, .
\end{align}
The first relation states that the T-matrix acts like an effective potential such that the full multi-scattering
 process is obtained in first Born approximation. The second equation
states how the T-matrix is obtained by iteration,
\begin{equation}\label{eq:scattering_expansion}
 \op{T} = \op{V} + \op{V} \op{G}_0 \op{V} + \op{V} \op{G}_0 \op{V} \op{G}_0 \op{V} + \ldots \,,
\end{equation}
implying that the multiple scattering events are categorized as single-, double-, \dots scattering events.

In our case the scattering potential $\op{V} = \sum_i \op{v}(i)$ is a sum of  single obstacle potentials
 differing merely in the position of the scatterer, and thus a
\emph{multi-scattering expansion} is appropriate. Let us define the corresponding single obstacle
t-matrices via
\begin{equation}
 \op{t}(i) = \op{v}(i) + \op{v}(i) \op{G}_0 \op{t}(i),
\end{equation}
describing multiple scattering form a single obstacle $i$.
 Then the  direct scattering expansion of Eq.~(\ref{eq:scattering_expansion}) can be reorganized into
\begin{align}\label{eq:multiple_scattering_expansion}
 \op{T} = \sum_{i} \op{t}_i + \sum_{i\neq j} \op{t}(i) \op{G}_0 \op{t}(j) +
\sum_{\substack{i\neq j \\ j \neq k}} \op{t}(i) \op{G}_0 \op{t}(j) \op{G}_0 \op{t}(k) + \ldots\, ,
\end{align}
where the first term accounts for the repeated scattering with the \emph{same} scatterer,
the second one repeated scattering with one scatterer followed by a series of scattering events
with a \emph{different} scatterer. The third term continues this series including scattering from a
 third scatterer, which may coincide with the first one.
It is now clear how the term involving $n$ t-matrices is constructed: pick all sequences consisting
 of $n$ obstacles such that all neighboring pairs are distinct. These sequences are represented by
 multiplying the single obstacle t-matrices sandwiched with an unperturbed propagator. The multiple
 scattering expansion, Eq.~(\ref{eq:multiple_scattering_expansion}),
remains valid in the case of hard obstacles provided the single obstacle t-matrix is calculated via
the full propagator  of the single impurity problem analogous to Eq.~(\ref{eq:full_propagator}).

The  dynamics of the tracer depends on the details of its local environment. Performing an
ensemble average over different initial positions of the tracer and measuring only the
relative displacements reduces the task to computing the \emph{disorder-averaged}
propagator $\overline{\op{G}}$. By Eqs.~(\ref{eq:full_propagator},\ref{eq:definition_tmatrix}), this
 is achieved by equivalently
evaluating the average many obstacle T-matrix $\overline{\op{T}}$. The multiple scattering expansion
 shows that to first order it is sufficient to
determine the average single-obstacle $t$ matrix
\begin{align}\label{eq:tmatrix_density_expansion}
 \overline{\op{T}} = n L^d \, \,\bar{\op{t}} + \mathcal{O}(n^2)
\end{align}
 since the next terms involve at least two obstacles.  Furthermore the disorder average restores
 translational symmetry, implying $\overline{\op{T}}$ is diagonal in a plane wave basis.

The correlations induced by the interaction are conventionally represented in terms of a self-energy
 $\op{\Sigma}$, defined via Dyson's equation
\begin{equation}\label{eq:Dyson}
 \overline{\op{G}} = \op{G}_0 + \op{G}_0 \,\op{\Sigma} \,\overline{\op{G}}  \, .
\end{equation}
Comparison with Eqs.~\eqref{eq:full_propagator}, \eqref{eq:tmatrix_density_expansion} shows that
 to first order in the obstacle density
\begin{equation}\label{eq:self-energy}
 \op{\Sigma} = n L^d \, \,\bar{\op{t}} + \mathcal{O}(n^2)\, .
\end{equation}
For the evaluation of the first-order correction in the density, it is thus sufficient to solve
 the dynamics of the trace for a single scatterer.

\section{A single scatterer}\label{sec:single}

\noindent
The motion of a single particle diffusing in the presence of a fixed spherical hard obstacle is
 exactly solvable which is the main result of this section.
 Here we use an approach which  uses a mixed real-space momentum-space representation, making the calculation
much more transparent. Second we choose a formulation reminiscent of  the scattering
 problem of a quantum particle of a hard disk using the Hilbert space of the free motion as reference. The solution of the
quantum scattering problem can be found in many textbooks, see for example Ref.~\citealp{Sakurai:Modern_Quantum_Mechanics}.
Then to account for the exclusion by the obstacle a fictitious dynamics of ghost particles is introduced,
such that the full propagator is a superposition of the obstructed motion and a free diffusion pole. The hard-sphere scattering
also arises in the context of the low-density expansion of  hard-sphere suspensions, solved
by Ackerson and Fleishman~\cite{Ackerson:1982} in real space and later by Felderhof and Jones~\cite{Felderhof:1983a,Felderhof:1983b,Jones:1984}.

Here we calculate the t-matrix for the single obstacle problem  via a partial wave decomposition of the solution of the Helmholtz equation.
To indicate that the operators evaluated here refer to the single impurity problem, we use small letters for the operators.
First, it is useful to consider  the unperturbed  Green function $\op{g}_0$ (which is of course identical to $\op{G}_0$) satisfying
\begin{equation}
 (-\text{i} \omega- D_0\nabla^2) \langle \vec{r} | \op{g}_0 | \vec{R} \rangle =  \delta(\vec{r}-\vec{R}) \, ,
\end{equation}
without restrictions on the values of the initial and terminal positions $\vec{r}$ and $\vec{R}$.
The propagator $\op{g}$ for the Brownian particle   moving in a two-dimensional plane in the presence of a single hard core obstacle of radius
$\sigma$ placed at the origin satisfies the equation of motion
\begin{equation}
 (-\text{i} \omega - D_0 \nabla^2) \langle \vec{r} | \op{g} | \vec{R} \rangle =  \delta(\vec{r}-\vec{R}) \vartheta(|\vec{R}|-\sigma)\, .
\end{equation}
Here the Heaviside step function $\vartheta(|\vec{R}|-\sigma)$ accounts for the fact that the tracer cannot start
 within the obstacle implying that the matrix elements vanish
 $\langle \vec{r} | \op{g} | \vec{R} \rangle = 0$ for $|\vec{R}| < \sigma$. The hard-core repulsion imposes a no-flux boundary condition on the propagator,
\begin{equation}\label{eq:no_flux}
 \vec{r} \cdot \vec{\nabla} \langle \vec{r} | \op{g} | \vec{R} \rangle = 0 \qquad  \text{for} \quad |\vec{r}| = \sigma \, ,
\end{equation}
at the surface of the obstacle. To make contact with the formal scattering theory, the propagator $\op{g}$ is to be interpreted
as the resolvent of some time-evolution operator. In the case of a hard disk, the dynamics
 has been defined only for particles initially outside of the obstacle. Completing the description,
 we assume that particles positioned inside an obstacle at the beginning behave as ghost particles
diffusing freely without feeling the exclusion. The resolvent associated to this operator then consists of the propagator we seek plus a diffusive
pole with a residue corresponding to the packing fraction of the single obstacle on the plane.

The solution will be obtained by switching to a mixed coordinate-momentum representation
\begin{equation}
 \langle \vec{r} | \op{g} | \vec{q} \rangle = \frac{1}{\sqrt{A}} \int \diff^2 \vec{R} \,
\langle \vec{r} | \op{g} | \vec{R} \rangle \text{e}^{\text{i} \vec{q} \cdot \vec{R}}\, ,
\end{equation}
where the plane wave states  $|\vec{q} \rangle $ are defined by their overlap with the positional
 representation $\langle \vec{r} | \vec{q} \rangle = \text{e}^{\text{i} \vec{q} \cdot \vec{r}}/\sqrt{A}$.
 The wavenumbers $\vec{q}$ form a discrete set such that the positional representation
 obeys periodic boundary conditions in the plane $A=L^2$; furthermore they  are properly
 normalized $\langle \vec{k} |\vec{q} \rangle = \delta_{\vec{k},\vec{q}}$. Then the two Green functions satisfy
\begin{align}\label{eq:diffusion}
 (-\text{i} \omega - D_0\nabla^2) \langle \vec{r} | \op{g}_0 | \vec{q} \rangle &= \frac{1}{\sqrt{A}} \text{e}^{\text{i} \vec{q} \cdot \vec{r} }
 \, , \\
\label{eq:diffusion2}
 (-\text{i} \omega - D_0\nabla^2) \langle \vec{r} | \op{g} | \vec{q} \rangle &= \frac{1}{\sqrt{A}} \text{e}^{\text{i} \vec{q} \cdot \vec{r} } \, ,
\end{align}
where for $  \langle \vec{r} | \op{g} | \vec{q} \rangle$ satisfies the no-flux
 boundary condition, Eq. (\ref{eq:no_flux}), for $|\vec{r}| > \sigma$ and describes the
 ghost particle for $|\vec{r}| < \sigma$.
The free motion allows for a plane wave solution
\begin{equation}
 \langle \vec{r} | \op{g}_0 | \vec{q} \rangle =  \frac{\text{e}^{\text{i} \vec{q} \cdot \vec{r}}/\sqrt{A} }{-\text{i} \omega + D_0 q^2} \, ,
\end{equation}
which also is the solution for $ \langle \vec{r} | \op{g} | \vec{q} \rangle$ for terminal position inside the obstacle.
From the  observation  that the   difference of Eqs. \eqref{eq:diffusion} and \eqref{eq:diffusion2},
\begin{equation}\label{eq:Helmholtz}
 (\text{i} \omega + D_0\nabla^2) \langle \vec{r} | \op{g}-\op{g}_0 | \vec{q} \rangle = 0\, ,
\end{equation}
vanishes for terminal positions outside the obstacle, the remaining case $|\vec{r}| > \sigma$ follows immediately.
The general solution decaying rapidly at infinity of  the two-dimensional source-free Helmholtz equation (\ref{eq:Helmholtz}) in polar
 coordinates  is given in terms of modified Bessel functions of the second kind $K_\nu(\cdot)$ as
\begin{equation}
\langle \vec{r} | \op{g}-\op{g}_0 | \vec{q} \rangle = \sum_{m=-\infty}^\infty a_m K_m(\mu r) \text{e}^{\text{i} m \varphi}\, ,
\end{equation}
where we abbreviated $\mu = \sqrt{-\text{i} \omega/D_0} =(1-i)\sqrt{\omega/2D_0}$ and $\varphi = \angle (\vec{r},\vec{q})$.
The coefficients $a_m$ are determined by the boundary condition $\vec{r} \cdot \vec{\nabla} \langle \vec{r} | \op{g} | \vec{q} \rangle = 0$. Noting
 that $\langle \vec{r} | \op{g}_0 | \vec{q} \rangle $ can be represented  using the
 Jacobi-Anger expansion \cite[Eq.~9.1.41]{Abramowitz:Handbook_of_Mathematical_Functions}
\begin{equation}
 \text{e}^{\text{i} z \cos \varphi} = \sum_{n=-\infty}^\infty \text{i}^n\, J_n(z) \text{e}^{\text{i} n \varphi} \, ,
\end{equation}
one obtains
\begin{equation}
a_m = \frac{- 1/\sqrt{A}}{-\text{i} \omega + D_0 q^2} \text{i}^m \frac{q J_m'(q\sigma)}{\mu K_m'(\mu \sigma)}.
\end{equation}

The propagator can now be evaluated in the momentum representation via Fourier transform.
 Here we shall need only the forward scattering matrix element
\begin{align}
\langle \vec{q} | \op{g} -\op{g}_0 | \vec{q} \rangle
&= \frac{1}{\sqrt{A} } \int_{|\vec{r}|> \sigma} \! \diff^2 \vec{r}\,
 \text{e}^{-\text{i} \vec{q} \cdot \vec{r}} \langle \vec{r} | \op{g}-\op{g}_0 | \vec{q} \rangle  \nonumber \\
\begin{split}
&= \frac{-2 \pi/A}{-\text{i} \omega + D_0 q^2} \sum_{m=-\infty}^\infty \frac{q J_m'(q\sigma)}{\mu K_m'(\mu \sigma)} \\
& \qquad \times \int_\sigma^\infty r \diff r \, J_m(q r) K_m(\mu r) \, ,
\end{split}
\end{align}
where we used the fact that for terminal position inside the obstacle, the
 free and full propagator are identical. The second line is obtained using again
 the Jacobi-Anger expansion
and integrating over the relative angle. With the help of the indefinite integral
\begin{align}
& \int r\diff r J_m(q r) K_m(\mu r) = \nonumber \\
& - \frac{r}{q^2+ \mu^2} \left[q K_m(\mu r) J_m'(q r)-\mu K_m'(\mu r) J_m(q r) \right] \, ,
\end{align}
and the relation \cite[Eq.~9.1.76]{Abramowitz:Handbook_of_Mathematical_Functions})
\begin{equation}
 \sum_{m=-\infty}^\infty J_m(z)^2 = 1 \, \Rightarrow \sum_{m=-\infty}^\infty J_m(z) J_m'(z) = 0 \, ,
\end{equation}
one  obtains the forward scattering amplitude in closed form
\begin{multline}
\langle \vec{q} | \op{g} -\op{g}_0 | \vec{q} \rangle = \\
-\frac{2 \pi\sigma^2/A}{\left(-\text{i} \omega + D_0 q^2\right)^2} \, D_0 q^2
\sum_{m=-\infty}^\infty J_m'(q \sigma)^2 \frac{K_m(\mu \sigma)}{\mu \sigma K_m'(\mu \sigma)}  \, .
\end{multline}
The t-matrix for the current case is again defined via the operator relation
$ \op{g} = \op{g}_0 + \op{g}_0 \op{t} \op{g}_0$,
and since the unperturbed propagator is diagonal in the wave number representation
\begin{equation}
 \langle \vec{k} | \op{g}_0 | \vec{q} \rangle = \frac{1}{-\text{i} \omega + D_0 q^2 } \delta_{\vec{k},\vec{q}}\, ,
\end{equation}
the t-matrix for forward scattering is readily obtained
\begin{align}
\langle \vec{q} | \op{t} | \vec{q} \rangle &=
-\frac{2 \pi\sigma^2}{A} D_0 q^2 \sum_{m=-\infty}^\infty J_m'(q \sigma)^2 \frac{K_m(\mu \sigma)}{\mu \sigma K_m'(\mu \sigma)} \, .
\end{align}
Note that the forward scattering amplitude is independent of the position of the scatterer.
The self-energy $\op{\Sigma}$ is to first order in the density, $n^* = \sigma^2 N/A$, equal to
 the average T-matrix of the single scattering t-matrices, Eq.~(\ref{eq:self-energy}),
\begin{align}
\langle \vec{q} | \op{\Sigma} | \vec{q} \rangle  &=
- 2\pi n^* D_0 q^2 \sum_{m=-\infty}^\infty J_m'(q \sigma)^2 \frac{K_m(\mu \sigma)}{\mu \sigma K_m'(\mu \sigma)} + \mathcal{O}(n^{*2})\, ,
\end{align}
which includes the motion of the ghost particles with residue $n^* \pi$. Since
 averaging over the disorder restores translational symmetry the self-energy is diagonal in the
wave number representation.
The average propagator then reads
\begin{multline} \label{eq:Sigma_result}
 \langle \vec{q} | \overline{\op{G}} | \vec{q} \rangle
= \Bigg[ \left(-\text{i} \omega + D_0 q^2\right)  \\
+ 2\pi n^* D_0 q^2
\sum_{m=-\infty}^\infty J_m'(q \sigma)^2 \frac{K_m(\mu \sigma)}{\mu \sigma K_m'(\mu \sigma)}
 \Bigg]^{-1}
\end{multline}
up to order $\mathcal{O}(n^*)$.
This form still contains the diffusive motion of a the ghost particle. Yet subtracting the
 corresponding pole yields up to first order in the packing fraction only $(1-n^* \pi)$ as common prefactor,
which reflects the total probability for a randomly placed tracer not overlap with an obstacle. Dropping this factor again, one can take the preceding result as the
conditional propagator for particles initially in the void space.

For reference we give also the corresponding result in three dimensions, $n^*_{3d} =  N \sigma^3/L^3$:
\begin{align}
 \langle \vec{q} | \overline{\op{G}} | \vec{q} \rangle
=& \Big[ (-\text{i} \omega + D_0 q^2)  \nonumber \\
&+ 4 \pi n^*_{3d} D_0 q^2 \sum_{\ell=0}^\infty  (2\ell+1)  \frac{[j_\ell'(q \sigma)]^2}{\mu \sigma k_\ell'(\mu \sigma) } k_\ell(\mu \sigma)  \Big]^{-1}\, .
\end{align}

\section{Velocity autocorrelation function}\label{sec:vacf}

\noindent
The simplest quantity characterizing deviations from simple diffusion is the averaged
 mean-square displacement $\delta r^2(t)$ or the velocity autocorrelation function
defined via
\begin{equation}
 Z(t) = \frac{1}{2 d} \frac{\diff^2}{\diff t^2} \delta r^2(t)\, , \qquad t>0 \, .
\end{equation}
The corresponding one-sided Fourier transform $Z(\omega) = \int_0^\infty \text{e}^{\text{i} \omega t} Z(t) \diff t$ is related
to the long-wavelength limit of the  Green function $G(q,\omega) := \langle \vec{q} | \overline{\op{G}} | \vec{q} \rangle$,
see e.g.~\cite{BoonYip:1980}. For completeness the derivation is repeated here. Since the mean-square displacement
is the second moment of the averaged real-space propagator, the VACF can be represented as
\begin{align}
 Z(t) = \frac{1}{2d} \frac{\diff^2}{\diff t^2} \int  \, r^2 \overline{P(\vec{r}t|\vec{0} 0)} \diff^d \vec{r} \, .
\end{align}
The averaged intermediate scattering function
\begin{equation}
 F(q,t) := \int \text{e}^{\text{i} \vec{q} \cdot \vec{r}}  \overline{P(\vec{r}t|\vec{0} 0)} \diff^d \vec{r} \, ,
\end{equation}
exhibits a  long wavelength expansion $F(q,t) = 1 - q^2 \delta r^2(t)/2d + \mathcal{O}(q^4)$, implying
\begin{equation}
 Z(t) = - \lim_{q\to 0} \frac{1}{q^2} \partial_t^2 F(q,t) \, .
\end{equation}
Since $G(q,\omega)$ corresponds to the  one-sided Fourier transform of $F(q,t)$, this last relation can be translated to the frequency domain as
\begin{equation}
\Real Z(\omega) =   \lim_{q\to 0} \frac{\omega^2}{q^2} \Real G(q,\omega) \, ,
\end{equation}
and
to the self-energy $\Sigma(q,\omega) = \langle \vec{q} | \op{\Sigma} | \vec{q} \rangle$ via
\begin{align}
 Z(\omega)  =& D_0 - \lim_{q\to 0} \frac{1}{q^2}  \Sigma(q,\omega) \, .
\end{align}
Specializing again to the planar case, Eq.~\eqref{eq:Sigma_result}, only the terms
$m=\pm 1$ contribute and we obtain to lowest order in the density
\begin{align}
 Z(\omega) &=  D_0+ \pi n^* D_0 \frac{K_1(\mu \sigma)}{\mu \sigma K_1'(\mu \sigma)} \, ,
\end{align}
where the frequency dependence is encoded in $\mu = (1-i)\sqrt{\omega/2D_0}$.
For reference let us provide also the corresponding result for the three-dimensional case
\begin{align}
 Z(\omega) &= D_0 + \frac{4\pi}{3} n^*_{3d} D_0 \frac{k_1(\mu \sigma)}{\mu \sigma k_1'(\mu \sigma)} \, .
\end{align}
 The velocity autocorrelation inherits the dynamic correlations due to the frozen
 landscape of obstacles, in particular, it displays non-analytic behavior for low-frequencies
\begin{multline}
Z(\omega) =  D_0+ \pi n^* D_0 \left[ -1  - \frac{\text{i} \omega \sigma^2}{2 D_0}
\ln \left( \frac{4 D_0}{-\text{i} \omega \sigma^2 } \right)
+ \frac{\text{i} \omega \sigma^2 \gamma}{D_0} \right] \\
+ \mathcal{O}(\omega^2 \ln^2 \omega) \, ,
\label{eq:VACF(omega)}
\end{multline}
where $\gamma =0.577\ldots$ denotes the Euler-Mascheroni constant. By a Green-Kubo relation, the zero-frequency limit yields
 the long-time diffusion constant $D$ in the presence of obstacles,
\begin{equation}
D = Z(\omega =0) = D_0(1-\pi n^*) + \mathcal{O}(n^{*2}),
\label{eq:long-time_diffusion}
\end{equation}
highlighting the suppression of transport due to the excluded volume.
A Fourier backtransform
\begin{equation}
 Z(t) = \frac{2}{\pi} \int_0^\infty \left[\Real Z(\omega) \right] \cos(\omega t)\diff \omega \, ,
\end{equation}
shows that the non-analytic low-frequency expansion corresponds
to  a long-time anomaly,
\begin{equation}
 Z(t) \simeq -\frac{\pi n^* \sigma^2 }{2 t^2}  \qquad \text{as} \quad t\to \infty \, ,
\end{equation}
with the same power-law as for the ballistic planar Lorentz model. Here
 the prefactor is first order in the density reflecting the fact that the tracer can encounter  the same obstacle
by diffusion many times without going through a series of scattering events with other frozen obstacles.

The diffusive motion at high frequencies  is singular too,
\begin{equation}
Z(\omega) =  D_0 \left(1 -\pi n^* \sqrt{\frac{D_0}{-\text{i} \omega \sigma^2}}\right) + \mathcal{O}(\omega^{-1}) \, ,
\end{equation}
which is reminiscent of the skin effect for electromagnetic waves  on a metal.
The corresponding skin penetration length $\delta := \sqrt{2D_0/\omega}$ is the characteristic length scale for
diffusive transport, and the reduction reflects the probability to encounter an obstacle within that length.
A Fourier back-transform yields the singular short-time behavior
\begin{equation}
 Z(t) = -  n^* D_0\sqrt{\frac{\pi D_0}{\sigma^2 t} } \, , \qquad t\to 0 \, ,
 \label{eq:st}
\end{equation}
characteristic for Brownian motion in an environment of hard obstacles.

\section{Simulation results and Discussion}\label{sec:simulation}

\noindent
We have performed Brownian dynamics simulation which test our analytic result for the VACF
and explore the range of validity of  the first order approximation to the low-density approximation.

The array of immobilized obstacles is generated by placing randomly $N$ hard disks of radius $\sigma$ in a
 plane of size $A=L^2$. The positions of the scatterers are independently drawn from a uniform distribution and scatterers may overlap which
in principle occurs at any density $n=N/A$. To minimize finite-size effects periodic boundary conditions are employed,
with typical system sizes of $L/\sigma = 10^4$.

A single tracer explores the void space of the frozen landscape by Brownian motion. Special care has to be taken to account
 for the hard-core exclusion at the obstacles.
Here we have relied on
an event-driven algorithm developed for hard sphere liquids~\cite{Scala:2007} recently employed also for
the three-dimensional Lorentz model close to the percolation transition~\cite{Lorentz_JCP:2008}. The basic idea is
to compute
 ballistic trajectories including the collisions with obstacles, which are interrupted by fictitious collisions with a solvent
 acting as a heat bath. In its simplest version, these kicks from the solvent are instantaneous at regular time intervals with
 period $\tau_B$ where a new velocity is drawn at random from a two-dimensional Maxwell distribution of variance $v^2$. At time
scales large with respect to the algorithmic time $\tau_B$, the  motion is Brownian and solves the diffusion
equation with a short-time diffusion constant $D_0 = v^2 \tau_B/4$. By construction, the tracer never leaves the void space and the hard-core
 repulsion is manifested merely in the usual specular scattering at the surface of the obstacles.

\begin{figure}[t]
\includegraphics[width=\linewidth]{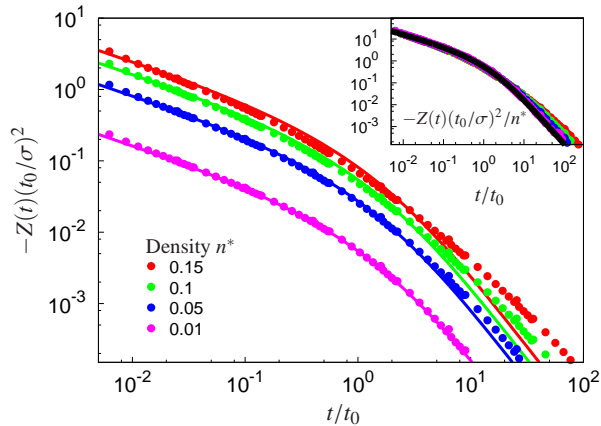}
\caption{Dimensionless negative VACF $-Z(t)$ for the dilute planar Lorentz gas for Brownian dynamics. Symbols
 correspond to  simulation results, full lines to the first-order density approximation. Density increases from
 bottom to top. Inset: Rescaling of $-Z(t)$ over~$n^{*}$}
\label{fig:vacf_inset}
\end{figure}
The simulations for Brownian tracer particles in the low-density range include  5  trajectories
 for each of the 155 different obstacle realizations drawn for each density, except for $n^{*}= 0.01$ where 500 different
 obstacle realizations where examined. We measure time in terms of microscopic scale $t_0 := \sigma^2/D_0$, i.e., the time needed for the particle
to diffuse  one obstacle radius without obstruction. The algorithmic time $\tau_B$ should be much smaller than $t_0$ and here we used $\tau_B = 0.0025 t_0$.
The negative velocity autocorrelation function $-Z(t)$ is displayed in Fig.~\ref{fig:vacf_inset} on double-logarithmic
 scales covering four non-trivial decades in time
and more than three orders of magnitude in signal.
For the smallest density $n^{*}=0.01$ the data coincide with the theoretical first-order density
approximation for all times. The time axis in the figure starts at $t=\tau_{B}$ and the small increase in the first
 data points visible is still affected by the algorithmic resolution. The curves corresponding to moderate densities still follow
the first-order low-density prediction at short times but start to deviate at long times. The long-time decay is  slower than the expected one
quantified by an apparent exponent which becomes smaller upon increasing the obstacle density. The crossover regime shows only a slight flattening
of the curvature  consistent with the observed increase of the exponent.
The inset in Fig.~\ref{fig:vacf_inset} corroborates very nicely the direct density dependence of the
 VACF following from Eq.~(\ref{eq:VACF(omega)}). All simulation
results overlap in the rectification with the  theoretical curve besides the exponent variations discussed above.
Let us mention that for the ballistic case
a numerical confirmation of the long-time tail with universal exponent has been achieved only recently~\cite{Lorentz_LTT:2007}, yet the amplitude for the
power law still deviated by 25\% from the theoretical value for $n^*=0.005$. Furthermore our simulation for a Brownian walker allows to compare the VACF for all times
to the theory since the VACF is known beyond its asymptotic behavior.

\begin{figure}[tb]
\includegraphics[width=\linewidth]{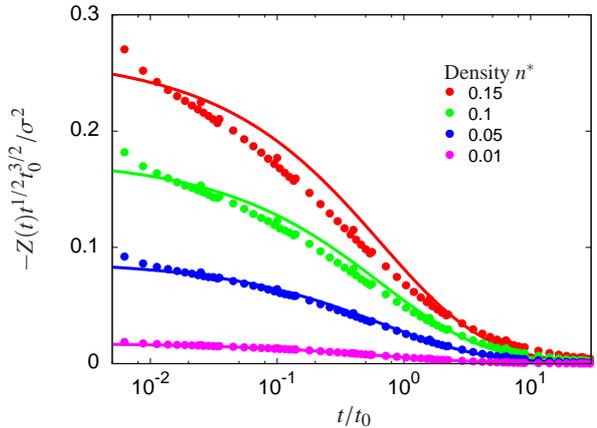}
\caption{Negative VACF $-Z(t)$ rectified by the short-time prediction $(t/t_{0})^{-1/2}$. Symbols correspond to our simulation results,
full lines to the first-order-density
approximation. Density increases from bottom to top.}
\label{fig:vacf_st}
\end{figure}
For a more thorough examination of the short-time behavior (Fig. \ref{fig:vacf_st}) it is advantageous to rectify the VACF
 with respect to the predicted $t^{-1/2}$ behavior of Eq.~(\ref{eq:st}). For the smallest density again a perfect agreement
 is observed. The moderate densities are still well described by the first-order-density theory
reflecting the fact that the particle did not have time to undergo collisions with more than one obstacle.
At short times the effects arising from a finite $\tau_{B}$ amplify as the system becomes denser, because
the number of collisions per $\tau_{B}$ is augmented by up to a factor of 15.
For the highest density $n=0.15$ we measure already 0.026
 collisions  on average between the algorithmic assignment of new random velocities. Assuming a Poissonian distribution for the scattering events
 the probability that two collisions
take place in this time interval is roughly about $0.1\%$ and therefore not completely negligible.

\begin{figure}[t]
\includegraphics[width=\linewidth]{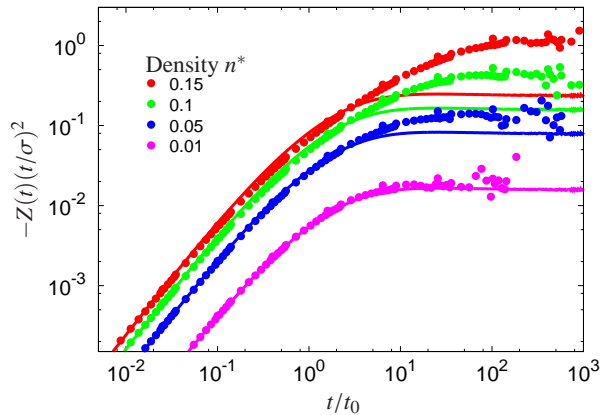}
\caption{Long-time behavior of $Z(t)$ rectified by the long-time tail $(t/t_{0})^{-2}$. Symbols correspond to the simulation results,
full lines are the first-order-density
approximation. Density increases from bottom to top.}
\label{fig:vacf_ltt}
\end{figure}

A rectification plot for the long-time behavior is depicted in Fig.~\ref{fig:vacf_ltt}.
 The VACF reaches the predicted scaling $t^{-2}$ for all densities, yet the observed plateau values at long times increase stronger than expected.
For the ballistic Lorentz gas it has been shown
 that such an increase arises due to a competition of
the  long-time tails and the critical relaxation at the percolation transition~\cite{Lorentz_LTT:2007}.
There, the underlying fractal structure induces anomalous transport for the
mean-square displacement~\cite{Lorentz_PRL:2006} resulting in a fractal power-law decay of the VACF.
The dynamics of the two-dimensional Brownian motion close to the percolation transition shall be discussed
elsewhere~\cite{Lorentz_2D:2010}.

\begin{figure}[t]
\includegraphics[width=\linewidth]{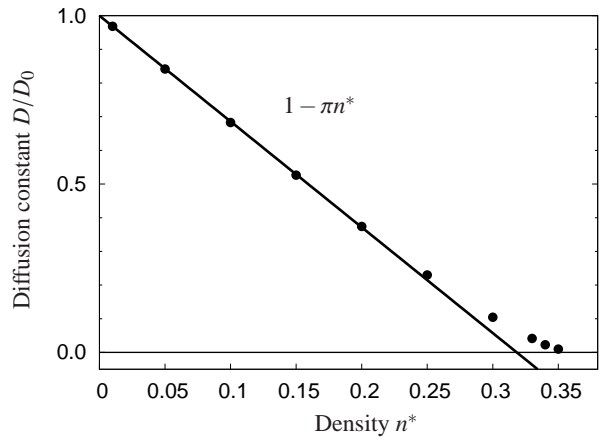}
\caption{Long-time diffusion coefficient $D$ as a function of reduced obstacle density $n^* = n \sigma^2$.
Symbols correspond the simulation results, the straight line $D/D_0 = 1- \pi n^* $ is the first-order-density approximation.}
\label{fig:diffusion}
\end{figure}

The long-time diffusion coefficients were extracted via $D=\lim_{t\to \infty} (1/4) \diff\delta r^{2}(t) /\diff t$ from
the simulated mean-square displacements. The suppression of the long-time diffusion coefficient predicted
 in Eq. (\ref{eq:long-time_diffusion}) is shown in Fig.~\ref{fig:diffusion}. Up to  densities of about $n^*= 0.2$ the data follow quite nicely
 the predictions; above, higher order corrections gain importance. Nevertheless the first-order theory provides a remarkably good description
 of the data for all densities.  Extrapolation suggests that the diffusion coefficient should vanish at $n^* = 1/\pi = 0.318..$ which is
surprisingly close to  the measured critical density of the percolation transition $n_c^* = 0.359..$~\cite{Lorentz_2D:2010}.

\section{Conclusion and Outlook}\label{sec:Conclusion}

\noindent
The notion that correlations quickly die out at time scales beyond some characteristic relaxation time of the problem has been shown
 to be incorrect in general. Besides
the meanwhile established long-time anomalies due to momentum conservation in fluids~\cite{Alder:1967,Alder:1970},
a second paradigm leading to persistent correlations is identified. Quenched disorder implies repeated encounters with the same obstacle, and
 the information encoded in the exclusion of the configuration space manifests itself in measurable quantities such as the mean-square displacement or
the velocity autocorrelation function.

 Previous studies focused on the ballistic motion in quenched disorder~\cite{Weijland:1968,Ernst:1971a} where
 the theoretical analysis is quite involved, and the long-time anomalies could be considered as merely a peculiarity of the Lorentz model.
The identification of similar persistent correlations for hopping transport in disordered lattices~\cite{Nieuwenhuizen:1987,Ernst:1987,Frenkel:1987}
suggests that memory effects may apply to a much larger class of systems. Here we have calculated the memory effects for a Brownian particle in a random environment
of hard scatterers
to first order in the obstacle density. We conclude that the only ingredient necessary for the long-time tails is the frozen disorder. Since disorder is
ubiquitous in nature and the effects arise at all  obstacle densities, we conclude that the power laws in the VACF are present for all real systems.

We have confirmed our analytical results by computer simulation for Brownian motion in a disordered array of obstacles. The Brownian particle
follows the first-order theory quantitatively at low densities and qualitatively for moderate ones confirming that the long-time anomaly persists for all densities where
diffusion occurs in the long-time limit. Furthermore, we have shown that the first-order-density expansion gives a reliable picture for Brownian dynamics,
 in contrast to the ballistic case.

The Brownian dynamics in the presence of hard obstacles displays a second power law in the VACF at short times which is due to single scattering events from an obstacle.
These effects have no analog in the ballistic case nor in hopping transport in disordered lattices.

In the present study the unobstructed dynamics is considered to be a random walk where momentum does not play a role. For a colloidal particle
 suspended in a fluid and moving
through a dilute course of obstacles, one may expect that both the vortex diffusion of momentum in the fluid as well as the repeated encounters with static
obstacles leads to an algebraic long-time decay of the VACF. Since the decay $t^{-d/2}$ due to vortex diffusion  for unconfined motion is slower than the one for
the obstructed motion $t^{-(d+2)/2}$ one may anticipate that momentum conservation is more important than steric hindrance at least for long times.
Yet, the obstacles can also carry away momentum
and the vortex diffusion in the fluid should be cut off at time scales where the agitated fluid encounters an obstacle. For the case of a wall in three dimensions
it has been shown that the long-time anomalies are suppressed~\cite{Felderhof:2005,Jeney:2008,Franosch:2009} to $t^{-5/2}$  rather than $t^{-3/2}$.
Hence, for a disordered system in a fluid
a competition between both scenarios appears to be relevant, yet no theory is available. Computer simulations for dense hard sphere liquids have indicated
a crossover from a positive tail due to vortex diffusion as well as a negative tail in the densely packed regime, where
the cages act as a quasi-frozen disordered environment~\cite{Williams:2006}. This phenomenology has been recently corroborated  for
 Lennard-Jones particles \cite{Glassy_GPU:2010}.

Hydrodynamic fluctuations in two-dimensional fluids lead to a series of divergences in the correlation functions. The diffusion coefficient in an infinite
system is expected to be (logarithmically) divergent since the fluid cannot carry momentum away fast enough. For a dilute density of static obstacles one may hope that
some of these effects are regularized in a natural way, giving an intriguing interplay of non-analytic behaviors; a theory for such a scenario, however, remains an open
question.

\section{Acknowledgments}

\noindent

Financial support  from the Deutsche Forschungsgemeinschaft via contract No.\ FR 850/6-1 is gratefully acknowledged.
This project is supported by the  German Excellence Initiative via the program ``Nanosystems Initiative Munich (NIM).''


\end{document}